\documentclass[12pt]{article}
\usepackage{epsfig}
\makeatletter
\@addtoreset{equation}{section}
\makeatother

\newcommand{\ret}{\nonumber\\}

\newcommand{\rbk}[1]{\left(#1\right)}
\newcommand{\sqbk}[1]{\left[#1\right]}

\newcommand{\bkt}[1]{\left\langle#1\right\rangle}

\renewcommand{\phi}{\varphi}
\newcommand{\ket}[1]{|#1\rangle}
\newcommand{\bra}[1]{\langle #1|}
\newcommand{\Tr}{{\rm Tr}}
\newcommand{\tbeta}{\tilde{\beta}}

\newcommand{\brap}[1]{\bra{\phi_{#1}}}
\newcommand{\brapp}[1]{\bra{\phi'_{#1}}}
\newcommand{\ketp}[1]{\ket{\phi_{#1}}}
\newcommand{\ketpp}[1]{\ket{\phi'_{#1}}}
\newcommand{\ketpi}[2]{\ket{\phi^{(#1)}_{#2}}}
\newcommand{\ketppi}[2]{\ket{\phi'^{(#1)}_{#2}}}
\newcommand{\brapi}[2]{\bra{\phi^{(#1)}_{#2}}}
\newcommand{\init}{_{{\rm init}}}
\newcommand{\fin}{_{{\rm fin}}}
\newcommand{\Hilb}{{\cal H}}
\newcommand{\bi}{{\bf i}}
\newcommand{\bj}{{\bf j}}
\newcommand{\DS}{\Delta S}
\begin{document}
\begin{flushright}
\small
technical note, cond-mat/0009244
\end{flushright}

\noindent
{\Large \bf Jarzynski Relations 
for Quantum Systems
and Some Applications}

\bigskip
\noindent
Hal Tasaki\footnote{
Department of Physics,
Gakushuin University,
Mejiro, Toshima-ku, Tokyo 171,
JAPAN

electronic address: hal.tasaki@gakushuin.ac.jp
}

\section{Introduction}
In a series of papers (see \cite{J1,J2,J3}
and references therein)
Jarzynski proved remarkable exact relations
in various dynamical systems
which relate nonequilibrium quantities
with equilibrium free energies.
Possible relations between 
Jarzynski's results
and the ``fluctuation theorem'' 
for steady state
nonequilibrium systems has 
been pointed out
by Crooks \cite{C}.
Piechocinska \cite{P} and Kurchan \cite{K} extended 
some of these results
to quantum systems.
See also \cite{Y} for a related (but different)
attempt at extending Jarzynski relations to 
quantum systems.

In this note,
we derive quantum analogues of 
Jarzynski's relations,
and discuss two applications, namely,
a derivation of the law of entropy increase
for general compound systems, and
a preliminary analysis of heat transfer between two
quantum systems at different temperatures.
We believe that the derivation of 
the law of entropy increase
in Section~\ref{s:S} is new and rather important.
(But the argument  
is essentially that in \cite{J2}.)

It is likely that many of the 
materials presented
here are known to experts in the field.
We nevertheless believe that to summarize
all these results in a self-contained 
(and, hopefully, transparent) manner 
can be useful for further investigation
in related (more difficult) problems.

We have tried to make
the present note technically self-contained.
The readers are suggested to go back to the
references for physical backgrounds.

\section{Jarzynski relations for quantum systems}
\label{s:J}
We concentrate here on a general quantum
mechanical system with a single
time-dependent Hamiltonian, 
and describe the derivation of 
quantum analogues of Jarzynski relations.
Some of the results are 
already derived
in \cite{P,K}.

\subsection{Setup}
Let us give abstract settings without specifying
physical interpretations.

Consider a quantum system with an \( N \)-dimensional
Hilbert space.
Take an arbitrary time-dependent Hamiltonian 
\( H(t) \)
in the time interval 
\( t\in[t\init,t\fin ] \),
and let \( H(t\init)=H \) and 
\( H(t\fin )=H' \).
For \( i=1,\ldots,N \),
we denote by 
\( \ket{\phi_{i}} \) and 
\( \ket{\phi'_{i}} \) the normalized eigenstates
of \( H \) and \( H' \), respectively,
with eigenvalues \( E_{i} \) and 
\( E'_{i} \).
Let \( U \) be the unitary operator
for time evolution during the whole 
period\footnote{
Let \( U(t) \) be the solution of 
the Schr\"{o}dinger 
equation
\( i\partial U(t)/\partial t=H(t)U(t) \)
with \( U(t\init)={\bf 1} \).
Then \( U=U(t\fin ) \).
}.

We assume that, at \( t=t\init \), the system is in the 
Gibbs state with inverse temperature 
\( \beta \),
which is described by the density matrix
\begin{equation}
    \rho\init=
    \frac{e^{-\beta H}}{Z(\beta)},
    \label{eq:rho1}
\end{equation}
with 
\( Z(\beta)=
\sum_{i=1}^N e^{-\beta E_{i}} \).
We denote by
\begin{equation}
    \bkt{\cdots}\init
    =
    \Tr[(\cdots)\rho\init],
    \label{eq:exp1}
\end{equation}
the expectation at \( t=t\init \), and by
\begin{equation}
    \bkt{\cdots}\fin
    =
    \Tr[(\cdots)U\rho\init U^{-1}],
    \label{eq:exp2}
\end{equation}
the expectation at \( t=t\fin  \).

Let us define a crucial quantity
\begin{equation}
    p_{i,j}=
    \frac{e^{-\beta E_{i}}}{Z(\beta)}
    \,
    |\brapp{j}U\ketp{i}|^2,
    \label{eq:pij}
\end{equation}
which (if one wishes) can be 
interpreted as the probability that one finds the 
system in the \( i \)-th eigenstate
of \( H \) at \( t=t\init  \)
and then in the \( j \)-th eigenstate
of \( H' \) at \( t=t\fin  \).
(See \cite{P,K} and footnote~\ref{fn:measure}.)
One can also think of \( p_{i,j} \) as 
a purely theoretical
quantity without giving it any interpretations.
Note that the unitarity (and the definition 
of \( Z(\beta) \))
guarantees the normalization
\( \sum_{i,j}p_{i,j}=1 \).
For any function \( f(E,E') \) of
two energy variables \( E \) and \( E' \),
we define its ``classical average'' by
\begin{equation}
    \overline{f(E,E')}
    =
    \sum_{i,j=1}^N
    p_{i,j}\,f(E_{i},E'_{j}).
    \label{eq:fbar}
\end{equation}
It is crucial to note that, in a general situation where
\( H \) and \( U^{-1}H'U \) do not commute, 
the classical average \( \overline{f(E,E')} \) does not
correspond to quantum mechanical expectation in any
obvious way.
However, for the simplest \( f \), we easily find
\begin{equation}
    \overline{E}=\bkt{H}\init ,\quad
    \overline{E'}=\bkt{H'}\fin ,
    \label{eq:EH}
\end{equation}
which will be crucial for the applications of the 
main equality.
\subsection{Jarzyinski's equality and inequality}
From the definitions (\ref{eq:fbar}) 
and (\ref{eq:pij}),
we find for any \( \tbeta \) that
\begin{eqnarray}
    \overline{e^{\beta E-\tbeta E'}}
    &=&
    \sum_{i,j=1}^N
    e^{\beta E_{i}-\tbeta E'_{j}}
    \,\frac{e^{-\beta E_{i}}}{Z(\beta)}
    \,
    |\brapp{j}U\ketp{i}|^2
    \ret
    &=&
    \frac{1}{Z(\beta)}
    \sum_{j=1}^Ne^{-\tbeta E'_{j}}
    \ret
    &=&
    \frac{Z'(\tbeta)}{Z(\beta)},
    \label{eq:Je1}
\end{eqnarray}
where we used the unitarity to get the second line, 
and introduced
\( Z'(\tbeta)=\sum_{j=1}^Ne^{-\tbeta E'_{j}} \).
We call (\ref{eq:Je1}) the Jarzynski equality.

By using the Jensen inequality 
\( \overline{\exp(f)}\ge \exp(\overline{f}) \),
we find that
\begin{equation}
    \exp\rbk{\beta \overline{E}
    -\tbeta \overline{E'}}
    \le
    \frac{Z'(\tbeta)}{Z(\beta)}.
    \label{eq:Ji1}
\end{equation}
By noting the identities (\ref{eq:EH}), 
this reduces to the following inequality for 
quantum mechanical expectation values.
\begin{equation}
    \beta\bkt{H}\init -\tbeta\bkt{H'}\fin 
    \le
    \log Z'(\tbeta)-\log Z(\beta),
    \label{eq:bH1}
\end{equation}
for any \( \tbeta \).

\bigskip
\noindent
{\bf Remark:}
(9/25/2000)
The inequality (\ref{eq:bH1}) (as well as other 
inequalities such as (\ref{eq:ebeZZ})
and (\ref{eq:DS>0}))
follows directly from the wellknown property of
the relative entropy as follows.
(We stress, however, that the Jarzynski's 
basic {\em equality}\/ contains much stronger
information than the corresponding inequality.)

Let \( \rho\init \) be an arbitrary density matrix
and let \( \rho\fin=U\rho\init U^{-1} \)
where \( U \) is an arbitrary unitary operator.
Then we have
\begin{equation}
    \Tr[\rho\init\log\rho\init]
    =\Tr[\rho\fin\log\rho\fin],
    \label{eq:vNinv}
\end{equation}
which is the well known invariance of the 
von Neuman entropy.
Let \( \rho' \) be an arbitrary density matrix.
Then the relative entropy \cite{W} satisfies
\begin{equation}
    S(\rho'|\rho\fin)
    =
    \Tr[\rho\fin(\log\rho\fin-\log\rho')]
    \ge0.
    \label{eq:relativeS}
\end{equation}
By combining (\ref{eq:vNinv}) and (\ref{eq:relativeS}),
we get
\begin{equation}
    \Tr[\rho\init\log\rho\init]\ge
    \Tr[\rho\fin\log\rho'].
    \label{eq:generalJ}
\end{equation}
This inequality reduces, for example,
to  (\ref{eq:ebeZZ}) by choosing both the
initial density matrix \( \rho\init \) and
the reference density matrix \( \rho' \) as 
Gibbs states.

\subsection{Basic 
applications to thermodynamics}
\label{ss:JS}
We now suppose that our quantum 
mechanical system is a macroscopic 
one, 
and imagine that the time dependent Hamiltonian
\( H(t) \) models an adiabatic operation
in thermodynamics.
(See, for example, \cite{Hal}.)
We further suppose that the Hamiltonian 
\( H(t) \) stays at \( H' \) sufficiently 
long time at the end of the operation,
so that the system reaches 
a macroscopic equilibrium.

By first setting \( \tbeta=\beta \),
we get an inequality for the 
(expectation value of the)
work \( W(U) \) done by the outside agent
(who controls the Hamiltonian)
to the system, i.e,
\begin{equation}
    W(U)\equiv\bkt{H'}\fin -\bkt{H}\init 
    \ge
    \frac{1}{\beta}\{
    \log Z(\beta)-\log Z'(\beta)
    \}.
    \label{eq:WU1}
\end{equation}
Since the relation of (the classical version of)
this inequality to the minimum work principle
is fully discussed in \cite{J1},
we do not repeat the discussion here.

Now in a thermodynamic system whose energy is
known to be \( U \),
its entropy can be obtained via 
a Legendre transformation as\footnote{
Throughout the present note,
we set the Boltzmann constant to unity.
}
\begin{equation}
    S(U)=\min_{\tbeta}\tbeta\{U-F(\tbeta)\}
    =\bar{\beta}\{U-F(\bar{\beta})\},
    \label{eq:SU}
\end{equation}
where \( F(\beta) \) is the 
Helmoholtz free energy,
and \( \bar{\beta} \) is the unique 
inverse temperature
at which the minimum is attained
(and which is nothing but the 
equilibrium value of
the inverse temperature).

From the basic inequality (\ref{eq:bH1}), we have
\begin{equation}
    \beta\{\bkt{H}\init -F(\beta)\}
    \le
    \tbeta\{\bkt{H'}\fin -F'(\tbeta)\}
    \label{eq:bHF1}
\end{equation}
for any \( \tbeta \),
where 
\( F(\beta)=-\beta^{-1}\log Z(\beta) \),
and 
\( F'(\tbeta)=-\tbeta^{-1}\log Z'(\tbeta) \).
From (\ref{eq:SU}), we see that the
left-hand side of (\ref{eq:bHF1})
is nothing but the entropy \( S \) of 
the initial state,
and the minimum over \( \tbeta \) of the
right-hand side is the entropy \( S' \)
of the final state.
Consequently we find from (\ref{eq:bHF1}) that
\begin{equation}
    S\le S',
    \label{eq:SS1}
\end{equation}
which is the law of entropy increase
in an adiabatic process\footnote{
The inequality (\ref{eq:bHF1}) is valid for
arbitrary finite systems.
But (\ref{eq:SS1}) is meaningful only for
macroscopic systems,
since the definition (\ref{eq:SU}) of entropy
should be used in the thermodynamic limit.
}.

\subsection{``Fluctuation theorem'' like symmetry}
\label{ss:Jf}
The reader interested in the topic of
Section~\ref{s:S} can skip the present
subsection.

Let a variable \( w \) take values of the form
\( E'_{j}-E_{i} \) for 
\( i,j=1,\ldots,N \), and
let
\begin{eqnarray}
    P_{U}(w)
    &=&
    \sum_{i,j=1}^N
    \chi[E'_{j}-E_{i}=w]\,p_{i,j}
    \ret
    &=&
    \sum_{i,j=1}^N
    \chi[E'_{j}-E_{i}=w]
    \,
    \frac{e^{-\beta E_{i}}}{Z(\beta)}
    \,
    |\brapp{j}U\ketp{i}|^2,
    \label{eq:PUw}
\end{eqnarray}
where the characteristic function is defined as
\( \chi[\mbox{true}]=1 \), and
\( \chi[\mbox{true}]=0 \).
This can be interpreted as the probability
that the difference of the measured
energies\footnote{
\label{fn:measure}
We suppose that, as in \cite{P,K},
one first measures \( H \) at
time \( t=t\init  \),
and the state contracts to the eigenstate 
which corresponds to the measured energy.
Then the system evolves according to \( U \),
and one measures \( H' \) at time 
\( t=t\fin  \).
} at the initial and the final
states is equal to \( w \).

Now an easy calculation shows
that 
\begin{eqnarray}
    e^{-\beta w}\,P_{U}(w)
    &=&
    \sum_{i,j=1}^N
    \chi[E'_{j}-E_{i}=w]
    \,
    e^{-\beta E'_{j}+\beta E_{i}}
    \,
    \frac{e^{-\beta E_{i}}}{Z(\beta)}
    \,
    |\brapp{j}U\ketp{i}|^2
    \ret
    &=&
    \frac{Z'(\beta)}{Z(\beta)}
    \sum_{i,j=1}^N
    \chi[E_{i}-E'_{j}=-w]
    \,
    \frac{e^{-\beta E'_{j}}}{Z'(\beta)}
    \,
    |\brap{i}U^{-1}\ketpp{j}|^2.
    \label{eq:ebwP}
\end{eqnarray}
With (\ref{eq:PUw}) in mind,
it is natural to define
\begin{equation}
    P_{U^{-1}}(w')
    =\sum_{i,j=1}^N
    \chi[E_{j}-E'_{i}=w']
    \,
    \frac{e^{-\beta E'_{i}}}{Z'(\beta)}
    \,
    |\brap{j}U^{-1}\ketpp{i}|^2.
    \label{eq:PUinv}
\end{equation}
This correspond to the inverse situation where the
initial state is the Gibbs state for 
\( H' \),
and the time evolution is given by 
\( U^{-1} \).
Then (\ref{eq:ebwP}) implies
\begin{equation}
    e^{-\beta w}\,P_{U}(w)=
    \frac{Z'(\beta)}{Z(\beta)}\,P_{U^{-1}}(-w).
    \label{eq:FT1}
\end{equation}

Consider a situation where \( H=H' \), and
the time evolution is symmetric in the 
sense that\footnote{
\label{fn:tr}
This is true when \( H(t) \) is symmetric in the
sense that 
\( H(t)=H(t\init +t\fin -t) \)
for any \( t \) such that 
\( t\init \le t\le t\fin  \),
and the matrix elements
\( \brap{i}H(t)\ketp{j} \) is real for
all \( t \) and \( i,j \).
}
\begin{equation}
    |\brap{j}U\ketp{i}|=|\brap{j}U^{-1}\ketp{i}|.
    \label{eq:UU}
\end{equation}
Then we have \( Z(\beta)=Z'(\beta) \),
and \( P_{U}(w)=P_{U^{-1}}(w) \).
Thus (\ref{eq:FT1}) take the 
significant form \cite{K}
\begin{equation}
    e^{-\beta w}P_{U}(w)=P_{U}(-w),
    \label{eq:}
\end{equation}
which (at least formally)
resembles the fluctuation theorem
for steady state\footnote{
One finds that the derivation is
quite similar (although the derivation
of the present result is rather trivial).
See especially \cite{M}.
}.
But it is questionable whether this equality
carries information about nonequilibrium systems
(like the original fluctuation theorem).
Later in Section~\ref{s:H}, we derive another
relation which also resembles 
the fluctuation theorem.
(See (\ref{eq:Pts2}).)
We do not go into details 
about the fluctuation theorem,
and just refer to recent works \cite{M,LS},
and those discuss possible relations between the
fluctuation theorem and the Jarzynski relations
\cite{C,J3}.

\section{The law of entropy increase 
for general compound
systems}
\label{s:S}
\subsection{Precise statement in thermodynamics}
The law of entropy increase represents the heart of
the second law of thermodynamics.
In the present note, we have already 
discussed the law of entropy 
increase (\ref{eq:SS1}),
but this is only for simple systems.

A {\em simple system}\/ is a basic 
notion in thermodynamics,
which stands for a system that can exchange
energy within it, and hence always  attains a uniform
temperature in equilibrium.
A {\em compound system}\/, on the other hand,
consists of several distinct simple systems which do 
not exchange energies with each other.
(In the usual language of thermodynamics,
they are separated by ``adiabatic walls.'')
An equilibrium state of a compound system is 
a composition of equilibrium states of 
each simple system that constitute it.

We assume that a thermodynamic 
(i.e., macroscopic)
equilibrium state of simple system is 
completely characterized by specifying its 
inverse temperature \( \beta \) and a 
set \( X \) of extensive variables\footnote{
This is not true at tricritical points, but
we do not go into such details here.
}.
In case of a fluid with \( M \) components in a 
container,
\( X=(V,N_{1},\ldots,N_{M}) \) where \( V \) is the
volume of the container and \( N_{i} \) the amount
of \( i \)-th substance.
We denote by \( (\beta;X) \) the equilibrium states
specified by \( \beta \) and \( X \).

Then an equilibrium state of a compound system 
can be written as 
\begin{equation}
    \{(\beta_{1};X_{1})|
    (\beta_{2};X_{2})|\cdots|
    (\beta_{n};X_{n})\},
    \label{eq:bX}
\end{equation}
where \( (\beta_{k};X_{k}) \) denotes the 
equilibrium state of the \( k \)-th simple system
constituting the whole compound system.
Here the vertical bars represent adiabatic walls.

Then the strongest form of 
the {\em law of entropy increase}\/
states that if an adiabatic operation
\begin{equation}
    \{(\beta_{1};X_{1})|\cdots|
    (\beta_{n};X_{n})\}
    \to
    \{(\beta'_{1};X'_{1})|\cdots|
    (\beta'_{m};X'_{m})\},
    \label{eq:a}
\end{equation}
is possible,
then one always has\footnote{
The {\em entropy principle}\/ states,
in addition to the law of entropy increase,
that the adiabatic operation (\ref{eq:a}) is
always possible when (\ref{eq:SScomp}) holds
(and when the conservations of masses are
satisfied).
We do not discuss this additional part of
the principle,
and focus only on the law of entropy increase.
}
\begin{equation}
    \sum_{k=1}^n 
    S(\beta_{k};X_{k})
    \le
    \sum_{\ell=1}^m 
    S(\beta'_{\ell};X'_{\ell}),
    \label{eq:SScomp}
\end{equation}
where \( S(\beta;X) \) is the entropy of a
simple system in state \( (\beta;X) \).
Note that we allowed the number of simple systems
to vary in the adiabatic operation 
(\ref{eq:a}), since 
we can insert or remove adiabatic walls during
an operation.

We stress that the law of entropy increase for
compound system is much stronger than that
for simple systems, and
(along with additivity of the entropy)
has far reaching implications.
See, for example, \cite{LY}.

\subsection{Statistical mechanical derivation}
We give a statistical mechanical derivation
of (\ref{eq:SScomp}).
We stress that what follows is an almost trivial 
and natural modification
of the argument in \cite{J2}.

We treat a model similar to Section~\ref{s:J} with
suitable modifications
necessary to represent compound systems.

We again model an adiabatic operation by a
time dependent Hamiltonian \( H(t) \)
with \( t \) in the range 
\( t\init\le t\le t\fin \).
We denote the initial Hamiltonian 
as \( H(t\init)=H \) and
the final Hamiltonian as
\( H(t\fin)=H' \).

We decompose the Hilbert space \( \Hilb \) of
the system as
\begin{equation}
    \Hilb=\bigotimes_{k=1}^n \Hilb_{k},
    \label{eq:Hilbk}
\end{equation}
and assume that the initial Hamiltonian has
the form
\begin{equation}
    H=\sum_{k=1}^n 
    {\bf 1}_{1}\otimes{\bf 1}_{2}
    \otimes\cdots\otimes
    {\bf 1}_{k-1}\otimes H_{k}
    \otimes{\bf 1}_{k+1}\otimes\cdots
    \otimes{\bf 1}_{n},
    \label{eq:HHk}
\end{equation}
where \( {\bf 1}_{k} \) is the identity operator
on \( \Hilb_{k} \),
and \( H_{k} \) acts only on \( \Hilb_{k} \).
For each \( k=1,\ldots,n \),
we denote by \( \ketpi{k}{i} \) with
\( i=1,\ldots,d_{k}\equiv\dim(\Hilb_{k}) \)
the normalized eigenstate of \( H_{k} \)
with the eigenvalue \( E^{(k)}_{i} \).
To represent states of the whole system,
we use a multi-index
\( \bi=(i_{1},i_{2},\ldots,i_{n}) \)
with \( i_{k}=1,\ldots,d_{k} \),
and define
\begin{equation}
    \ket{\Phi_{\bi}}=
    \bigotimes_{k=1}^n\ketpi{k}{i_{k}}.
    \label{eq:Phii}
\end{equation}

Similarly
we decompose 
the same Hilbert space as
\begin{equation}
    \Hilb=
    \bigotimes_{\ell=1}^m \Hilb'_{\ell},
    \label{eq:Hilbl}
\end{equation}
and assume that the final Hamiltonian has
the form
\begin{equation}
    H'=\sum_{\ell=1}^m 
    {\bf 1}'_{1}\otimes{\bf 1}'_{2}
    \otimes\cdots\otimes
    {\bf 1}'_{\ell-1}\otimes H'_{\ell}
    \otimes{\bf 1}'_{\ell+1}\otimes\cdots
    \otimes{\bf 1}'_{m},
    \label{eq:HHl}
\end{equation}
where \( {\bf 1}'_{\ell} \) is the identity operator
on \( \Hilb'_{\ell} \),
and \( H'_{\ell} \) acts only on 
\( \Hilb'_{\ell} \).
For each \( \ell=1,\ldots,m \),
we denote by \( \ketppi{\ell}{j} \) with
\( j=1,\ldots,d'_{\ell}
\equiv\dim(\Hilb'_{\ell}) \)
the normalized eigenstate of \( H'_{\ell} \)
with the eigenvalue \( E'^{(\ell)}_{j} \).
We again use a multi-index
\( \bj=(j_{1},j_{2},\ldots,j_{m}) \)
with \( j_{\ell}=1,\ldots,d'_{\ell} \),
and define
\begin{equation}
    \ket{\Phi'_{\bj}}=
    \bigotimes_{\ell=1}^m\ketppi{\ell}{j_{\ell}}.
    \label{eq:Phij}
\end{equation}

We assume that at \( t=t\init \) the 
\( k \)-th simple system is in the
Gibbs state with inverse temperature 
\( \beta_{k} \).
The whole state is then
represented by the density matrix
\begin{equation}
    \rho\init=
    \bigotimes_{k=1}^n
    \frac{e^{-\beta_{k}H_{k}}}
    {Z_{k}(\beta_{k})},
    \label{eq:rhoinitcomp}
\end{equation}
where 
\( Z_{k}(\beta)=\sum_{i=1}^{d_{k}}
e^{-\beta E_{i}^{(k)}}\).

Then the system evolves according to the
time dependent Hamiltonian \( H(t) \) 
until \( t=t\fin \).
We assume that, near the end of the
operation,
\( H(t) \) stays at \( H' \) for sufficiently 
long time so that each of the \( m \)
simple systems reach equilibrium.
(Note that they are not necessarily
described by the exact Gibbs states.
See, for example, the discussion at the end
of \cite{Hal}.)
We denote by \( U \) the
time evolution unitary operator
for the whole process,
and denote by
\begin{equation}
    \bkt{\cdots}\init=
    \Tr[(\cdots)\rho\init],
    \label{eq:expinit}
\end{equation}
the expectation at \( t=t\init \),
and by
\begin{equation}
    \bkt{\cdots}\fin=
    \Tr[(\cdots)U\rho\init U^{-1}],
    \label{eq:expfin}
\end{equation}
the expectation at \( t=t\fin \).

Following (\ref{eq:pij}), we define
\begin{equation}
    p_{\bi,\bj}=
    \rbk{\prod_{k=1}^n
    \frac{\exp[-\beta_{k}E_{i_{k}}^{(k)}]}
    {Z_{k}(\beta_{k})}}
    |\bra{\Phi'_{\bj}}U\ket{\Phi_{\bi}}|^2,
    \label{eq:pbij}
\end{equation}
and
\begin{equation}
    \overline{
    f(E^{(1)},\ldots,E^{(n)},
    E'^{(1)},\ldots,E'^{(m)})
    }
    =
    \sum_{\bi,\bj}
    p_{\bi,\bj}\,
    f(E^{(1)}_{i_{1}},\ldots,E^{(n)}_{i_{n}},
    E'^{(1)}_{j_{1}},\ldots,E'^{(m)}_{j_{m}}).
    \label{eq:fbarc}
\end{equation}
Then one easily finds for
any \( \tbeta_{1},\ldots,\tbeta_{m} \)
that
\begin{eqnarray}
    \overline{
    \exp[\sum_{k=1}^n\beta_{k}E^{(k)}
    -\sum_{\ell=1}^m\tbeta_{\ell}E'^{(\ell)}]
    }
    &=&
    \rbk{\prod_{k=1}^n
    \frac{1}{Z_{k}(\beta_{k})}}
    \sum_{\bi,\bj}
    \rbk{\prod_{\ell=1}^m
    \exp[-\tbeta_{\ell}E'^{(\ell)}_{j_{\ell}}]}
    |\bra{\Phi'_{\bj}}U\ket{\Phi_{\bi}}|^2
    \ret
    &=&
    \frac{\prod_{\ell=1}^mZ'_{\ell}(\tbeta_{\ell})}
    {\prod_{k=1}^nZ_{k}(\beta_{k})},
    \label{eq:ebZZ}
\end{eqnarray}
where 
\( Z'(\tbeta)=\sum_{j=1}^{d'_{\ell}}
e^{-\tbeta E'^{(\ell)}_{j}}\).
By using the Jensen inequality, we have
\begin{equation}
    \exp[\sum_{k=1}^n\beta_{k}\overline{E^{(k)}}
    -\sum_{\ell=1}^m
    \tbeta_{\ell}\overline{E'^{(\ell)}}]
    \le
    \frac{\prod_{\ell=1}^m
    Z'_{\ell}(\tbeta_{\ell})}
    {\prod_{k=1}^nZ_{k}(\beta_{k})}.
    \label{eq:ebeZZ}
\end{equation}
Noting that\footnote{
To be precise,
the left-hand side should be
\( \bkt{{\bf 1}_{1}\otimes
\cdots\otimes{\bf 1}_{k-1}
\otimes H_{k}\otimes{\bf 1}_{k+1}
\otimes\cdots
\otimes{\bf 1}_{n}}\init \).
}
\begin{equation}
    \bkt{H_{k}}\init
    =\overline{E^{(k)}},\quad
    \bkt{H'_{\ell}}\fin=
    \overline{E'^{(\ell)}},
    \label{eq:HEHE2}
\end{equation}
we finally get
\begin{equation}
    \sum_{k=1}^n\beta_{k}
    \{\bkt{H_{k}}\init
    -F_{k}(\beta_{k})\}
    \le
    \sum_{\ell=1}^m\beta_{\ell}
    \{\bkt{H'_{\ell}}\fin
    -F'_{\ell}(\tbeta_{\ell})\},
    \label{eq:HF2}
\end{equation}
for any 
\( \tbeta_{1},\ldots,\tbeta_{m} \),
where 
\( F_{k}(\beta)=
-\beta^{-1}\log Z_{k}(\beta) \)
and 
\( F'_{\ell}(\tbeta)=
-\tbeta^{-1}\log Z'_{\ell}(\tbeta) \).
Then the desired inequality
\begin{equation}
    \sum_{k=1}^nS_{k}
    \le
    \sum_{\ell=1}^mS'_{\ell},
    \label{eq:Sineq2}
\end{equation}
follows (for a macroscopic system)
as in Section~\ref{ss:JS}

\subsection{Remarks}
It is quite remarkable that 
Jarzynski's method provides us with such a simple
(indeed almost trivial) proof of 
the strongest form of the law of entropy principle.

It should be noted, however, 
that the present derivation works only when the
initial state is a product of exact Gibbs states.
The heart of modern statistical physics is
that macroscopic system in equilibrium can be
modelled by various different distributions,
and all of them give rise to the same
(true) thermodynamics.
According to this spirit, we must be
able to extend the present derivation to
a much wider class of initial states,
but it seems rather difficult for the moment.
Note that for simple systems,
there exists a proof of the second law
of thermodynamics which works
for a large class of distributions
\cite{Hal},
but the argument there can hardly be
applied to compound systems.

\section{Heat transfer 
between two quantum systems}
\label{s:H}
We briefly discuss preliminary results obtained
by applying the present techniques to a 
highly nonequilibrium transient phenomenon.
More precisely, we put two quantum system in
different temperatures into a thermal contact,
and wish to examine the heat flow from 
one system to the other.
This is essentially contained in Jarzynski's 
work \cite{J3},
but our motivation here seems less ambitious.

Consider a quantum mechanical system consisting of 
two systems whose Hilbert spaces are
\( \Hilb_{1} \) and \( \Hilb_{2} \).
Then the Hilbert space for the whole system
is \( \Hilb=\Hilb_{1}\otimes\Hilb_{2} \).
We assume that the Hamiltonian of the model
is time independent, and write it in the form
\begin{equation}
    H=
    H_{1}\otimes{\bf 1}_{2}
    +{\bf 1}_{1}\otimes H_{2}
    +H_{\rm int},
    \label{eq:Hint}
\end{equation}
where \( H_{k} \) acts
only on \( \Hilb_{k} \) (\( k=1,2 \)) while
\( H_{\rm int} \) acts on the
whole \( \Hilb \).
For \( k=1,2 \),
we denote by
\( \ketpi{k}{i} \) 
the normalized eigenstate of \( H_{k} \)
with the eigenvalue \( E^{(k)}_{i} \)
where \( i=1,\ldots,d_{k}\equiv\dim(\Hilb_{k}) \).

We assume that initially
(which corresponds to \( t=0 \))
two systems are independently in equilibrium
at inverse temperatures \( \beta_{1} \)
and \( \beta_{2} \), respectively.
This corresponds to the density matrix
\begin{equation}
    \rho\init=
    \frac{e^{-\beta_{1}H_{1}}}{Z_{1}(\beta_{1})}
    \otimes
    \frac{e^{-\beta_{2}H_{2}}}{Z_{2}(\beta_{2})},
    \label{eq:rhoinitHF}
\end{equation}
with 
\( Z_{k}(\beta)=\sum_{i=1}^{d_{k}}
e^{-\beta E^{(k)}_{i}}\).
Then the expectation at time \( t \)
is given by
\begin{equation}
    \bkt{\cdots}_{t}
    =
    \Tr[(\cdots)U(t)\rho\init U(-t)],
    \label{eq:bktt}
\end{equation}
where the unitary operator for time evolution 
is \( U(t)=e^{-iHt} \).

In order to test for the possible
heat transfer between the two systems,
we examine the quantity\footnote{
To be precise,
\( H_{1} \) and \( H_{2} \) should be
\( H_{1}\otimes{\bf 1}_{2} \) and
\( {\bf 1}_{1}\otimes H_{2} \),
respectively.
We make the same abbreviation in what
follows.
}
\begin{equation}
    \DS(t)=\beta_{1}\{
    \bkt{H_{1}}_{t}-\bkt{H_{1}}_{0}\}
    +
    \beta_{2}\{
    \bkt{H_{2}}_{t}-\bkt{H_{2}}_{0}\}.
    \label{eq:DS}
\end{equation}
Since 
\( \bkt{H_{k}}_{t}-\bkt{H_{k}}_{0} \)
is the increase of energy in the \( k \)-th
system,
\( \DS(t) \) can be interpreted as
the {\em increase of entropy}\/ in the
whole system, 
provided that the change in the inverse temperatures
of the two systems are small 
and the contribution of \( H_{\rm int} \)
to the total entropy is negligible.

To examine \( \DS(t) \),
we again introduce the probability
\begin{equation}
    p_{i,j;\ell,m}(t)
    =
    \frac{e^{-\beta_{1}E^{(1)}_{i}}}
    {Z_{1}(\beta_{1})}
    \frac{e^{-\beta_{2}E^{(2)}_{j}}}
    {Z_{2}(\beta_{2})}
    |\brapi{2}{m}\brapi{1}{\ell}U(t)
    \ketpi{1}{i}\ketpi{2}{j}|^2,
    \label{eq:pijlm}
\end{equation}
and the corresponding classical average
of a function
\( f(E^{(1)},E^{(2)},E'^{(1)},E'^{(2)}) \)
as
\begin{equation}
    [f(E^{(1)},E^{(2)},E'^{(1)},E'^{(2)})]_{t}
    =
    \sum_{i,j,\ell,m}
    p_{i,j;\ell,m}(t)\,
    f(E^{(1)}_{i},E^{(2)}_{j},
    E'^{(1)}_{\ell},E'^{(2)}_{m}),   
    \label{eq:ft}
\end{equation}
where we avoided the notation 
\( \overline{f} \) since the average
depends on \( t \).
Note again that 
\begin{equation}
    [E^{(k)}]_{t}=\bkt{H_{k}}_{0},\quad
    [E'^{(k)}]_{t}=\bkt{H_{k}}_{t}.
    \label{eq:EH3}
\end{equation}
Now it is automatic to check that
\begin{equation}
    \sqbk{
    e^{\beta_{1}(E^{(1)}-E'^{(1)})
    +\beta_{2}(E^{(2)}-E'^{(2)})
    }}_{t}=1,
    \label{eq:ebb1}
\end{equation}
for any \( t \).
Then the Jensen inequality
\( [e^f]_{t}\ge e^{[f]_{t}} \)
implies
\begin{equation}
    \DS(t)\ge0,
    \label{eq:DS>0}
\end{equation}
which shows the (wellknown) fact that heat never
flows from the colder to the hotter.
We stress that (\ref{eq:DS>0}) has been {\em proved
rigorously}\/ (and indeed very easily)
for quite general quantum systems.

To get a stronger result,
let 
\( s=\beta_{1}(E'^{(1)}-E^{(1)})
    +\beta_{2}(E'^{(2)}-E^{(2)}) \).
Then \( [s]_{t}=\DS(t) \).
Since (\ref{eq:ebb1}) says \( [e^{-s}]_{t}=1 \),
one has
\begin{equation}
    \DS(t)=[s]_{t}
    =[e^{-s}-1+s]_{t}
    =[d(s)]_{t},
    \label{eq:DSt}
\end{equation}
where
\begin{equation}
    d(x)=e^{-x}-1+x
    =\sum_{n=2}^\infty\frac{(-x)^n}{n!}
    \label{eq:dx}
\end{equation}
satisfies \( d(0)=0 \)
and \( d(x)>0 \) for any \( x\ne0 \).
(\ref{eq:DSt}) and (\ref{eq:ft}) 
lead to the expression
\begin{equation}
    \DS(t)=\sum_{i,j,\ell,m}
    p_{i,j,\ell,m}(t)\,
    d[\beta_{1}(E'^{(1)}_{\ell}-E^{(1)}_{i})
    +\beta_{2}(E'^{(2)}_{m}-E^{(2)}_{j})],
    \label{eq:DSfinal}
\end{equation}
which is remarkable in that 
every term in the sum is nonnegative.
This means that, when one tries to construct
a {\em rigorous lower bound}\/ for \( \DS(t) \)
(which would rigorously establish the 
existence of a finite heat flow),
one can freely throw away unwanted terms in 
(\ref{eq:DSfinal}), only keeping well controlled
terms in the sum.
(Moreover \( d(x) \) can be bounded from below by 
simpler functions if necessary.)
Although we have no concrete estimates in this
abstract setting, we hope one can construct
meaningful rigorous lower bounds for
\( \DS(t) \) by examining typical concrete
models.

Let us make one more
remark in connection with the
fluctuation theorem.
If we define
\begin{equation}
    P_{t}(s)=\sum_{i,j,\ell,m}
    p_{i,j,\ell,m}(t)
    \,
    \chi[\beta_{1}(E'^{(1)}_{\ell}-E^{(1)}_{i})
    +\beta_{2}(E'^{(2)}_{m}-E^{(2)}_{j})=s].   
    \label{eq:Pts}
\end{equation}
Then exactly as in Section~\ref{ss:Jf},
we can prove
\begin{equation}
    e^{-s}P_{t}(s)=P_{-t}(-s),
    \label{eq:Pts1}
\end{equation}
or, for models with time reversal 
symmetry\footnote{
In the sense that
\( |\brapi{2}{m}\brapi{1}{\ell}U(t)
    \ketpi{1}{i}\ketpi{2}{j}|=
    |\brapi{2}{m}\brapi{1}{\ell}U(-t)
    \ketpi{1}{i}\ketpi{2}{j}|\)
for any \( i,j,\ell.m \).
See footnote~\ref{fn:tr}.
}
\begin{equation}
    e^{-s}P_{t}(s)=P_{t}(-s),
    \label{eq:Pts2}
\end{equation}
which has the form of ``fluctuation theorem.''
This is a special case of 
Jarzynski's 
``detailed fluctuation theorem'' \cite{J3}.

\bigskip\bigskip\bigskip

It is a pleasure to thank Shin-ichi Sasa for 
introducing me to the whole subject, and for
indispensable discussions.


\end{document}